# A methane line list with sub-MHz accuracy in the 1250 to 1380 cm⁻¹ range from optical frequency comb Fourier transform spectroscopy

**Matthias Germann[1], Adrian Hjältén[1], Vincent Boudon[2], Cyril Richard[2], Karol Krzempek[3], Arkadiusz Hudzikowski[3], Aleksander Głuszek[3], Grzegorz Soboń[3] and Aleksandra Foltynowicz[1,+]**

[1]Department of Physics, Umeå University, 901 87 Umeå, Sweden

[2] Laboratoire Interdisciplinaire Carnot de Bourgogne, UMR 6303 CNRS/Université Bourgogne Franche-Comté, 9 Av. A. Savary, BP 47870, F-20178 Dijon Cedex, France

[3] Laser & Fiber Electronics Group, Faculty of Electronics, Photonics and Microsystems, Wrocław University of Science and Technology, Wybrzeże Wyspiańskiego 27, 50-370 Wrocław, Poland

⁺aleksandra.foltynowicz@umu.se

**Abstract**: We use a Fourier transform spectrometer based on a difference frequency generation optical frequency comb to measure high-resolution, low-pressure, room-temperature spectra of methane in the 1250 – 1380-cm⁻¹ range. From these spectra, we retrieve line positions and intensities of 678 lines of two isotopologues: 157 lines from the $^{12}CH_4$ $v_4$ fundamental band, 131 lines from the $^{13}CH_4$ $v_4$ fundamental band, as well as 390 lines from two $^{12}CH_4$ hot bands, $v_2 + v_4 - v_2$ and $2v_4 - v_4$. For another 165 lines from the $^{12}CH_4$ $v_4$ fundamental band we retrieve line positions only. The uncertainties of the line positions range from 0.19 to 2.3 MHz, and their median value is reduced by a factor of 18 and 59 compared to the previously available data for the $^{12}CH_4$ fundamental and hot bands, respectively, obtained from conventional FTIR absorption measurements. The new line positions are included in the global models of the spectrum of both methane isotopologues, and the fit residuals are reduced by a factor of 8 compared to previous absorption data, and 20 compared to emission data. The experimental line intensities have relative uncertainties in the range of 1.5 – 7.7%, similar to those in the previously available data; 235 new $^{12}CH_4$ line intensities are included in the global model.

## 1    Introduction

Methane, $CH_4$, plays a central role in both applied and fundamental sciences. A potent greenhouse gas, methane is of great interest to atmospheric physics [1-3]. Outside of Earth's atmosphere, methane has been found on several other planets of our solar system, such as Jupiter, Saturn, Uranus, or Neptune [4-7], as well as on their satellites Titan and Triton [8, 9], while its role on Mars is controversial [10]. Beyond our solar system, methane has been found on several exoplanets [11-14], as well as other celestial bodies such as brown dwarfs or comet tails (see Ref. [15] and references therein). Therefore, methane attracts interest from the astrophysics community. From the point of view of pure molecular physics, studying the methane molecule paved the way for our modern understanding of chemical bonding and molecular structure. Methane has, for example, been a testing ground for electronic structure calculations (see Ref. [16] and references therein). Furthermore, concepts as fundamental as nuclear spin symmetry conservation, or yet-to-be-detected tiny energy shifts due to the parity violating weak interaction have been studied or discussed by example of the methane molecule (see Refs [16, 17] and references therein).

Precise and accurate spectroscopic information about the methane molecule is paramount for numerous studies in these fields. However, despite the apparently simple molecular structure, the rotation-vibration spectrum of methane is complex. Being a tetrahedral molecule, methane has nine vibrational degrees of freedom that give rise to four normal modes of vibration: the nondegenerate symmetric stretch mode $\nu_1$, the doubly degenerate symmetric bend mode $\nu_2$, as well as two triply degenerate modes, the antisymmetric stretch $\nu_3$, and the antisymmetric bend $\nu_4$. The normal mode frequencies fulfill the approximate relation $\nu_1 \approx \nu_3 \approx 2\nu_2 \approx 2\nu_4$. Vibrational fundamental bands hence overlap with each other as well as with overtone, combination and hot bands. This results in a highly congested and irregular spectrum already at room temperature, and even more pronounced at higher temperatures [15, 16, 18, 19]. Therefore, accurate and precise spectroscopic data of a large number of $CH_4$ lines are needed to develop and validate theoretical models of the methane spectrum.

Different spectroscopic techniques have been used to study the rotation-vibration spectrum of methane across the near- and mid-infrared range, such as Fourier-transform infrared (FTIR) absorption spectroscopy [16, 18, 20-23], FTIR emission spectroscopy [24, 25], THz synchrotron measurements [26], laser absorption spectroscopy [27, 28], cavity ringdown-spectroscopy [29], sub-Doppler spectroscopy [30-32], and frequency comb spectroscopy [33-35]. However, the available data around 8 μm, a spectral region lying in the atmospheric water window, stem largely from FTIR absorption and emission spectroscopy with limited accuracy and precision. The most recent global analysis of the methane rotation-vibration spectrum by Amyay et al. [25, 36], is, for the fundamental transitions around 8 μm, to a considerable part still based on conventional FTIR data recorded by Champion et al. [21] and Brown et al. [20] nearly 40 years ago. The precision of the line positions from these data is estimated to be 1.8 MHz, and the 'absolute accuracy' to be 3 MHz [21]. The precision of the line intensities ranges from 1.5% to 10% with an 'absolute accuracy' of 3% [20]. The HITRAN2020 database states considerably more conservative uncertainty estimates, 30 to 300 MHz for line positions and at least 20% for line intensities of the $\nu_4$ fundamental band [37].

With fully-stabilized optical frequency comb sources and comb-based spectrometers becoming available in the 8 μm range [38, 39], considerably higher frequency accuracy and precision can be reached. Comb-based Fourier transform spectroscopy lends itself particularly useful to studying the congested and irregular spectrum of methane, since this technique can acquire high-resolution spectra simultaneously over large spectral intervals, and eliminates the need to probe each absorption line individually by tuning the laser source [40]. Recently, Hjältén et al. [41] used optical frequency comb Fourier transform spectroscopy to record high-resolution spectra of the $\nu_1$ fundamental band and the $\nu_1 + \nu_2 - \nu_2$ hot band of nitrous oxide ($N_2O$), achieving line positions with uncertainties of the order of 100 kHz. The results were in good agreement with the earlier study by AlSaif et al. [42, 43], who recorded the $\nu_1$ fundamental band using a comb-referenced continuous-wave (CW) quantum cascade laser. While dual-comb spectroscopy with quantum-cascade lasers has been successfully applied to detect methane in the 8 μm spectral range [44-46], this method still relies on absorption lines with known frequencies for frequency calibration [44, 47].

Here we take advantage of the high resolution, absolute frequency calibration, large bandwidth and high spectral brightness provided by the previously developed optical frequency comb Fourier transform spectrometer (FTS) [41] to record methane absorption spectra in the range from 1250 to 1380 cm$^{-1}$. From these spectra, we retrieve line positions and intensities of over 650 absorption lines from the $\nu_4$ fundamental bands of $^{12}CH_4$ and $^{13}CH_4$, as well as from two $^{12}CH_4$ hot bands, $\nu_2 + \nu_4 - \nu_2$ and $2\nu_4 - \nu_4$. For another 165 lines from the $^{12}CH_4$ fundamental band, we retrieve line positions only. The typical (i.e., median) uncertainties are 430 kHz for line positions and 3% for line intensities. We use the line positions and intensities to replace and/or complement the data previously used in the global model of the methane absorption spectrum. We achieve significant improvements in the residuals of the line position fits of the effective Hamiltonian.

## 2   Experimental setup and procedures

The comb-based Fourier transform spectrometer has been described in detail previously [41]. In brief, it consists of three main building blocks: a mid-infrared (MIR) optical frequency comb source, a multi-pass absorption cell, and an FTS.

The MIR optical frequency comb was generated from two near-infrared femtosecond pulse trains via difference frequency generation in an orientation-patterned gallium phosphide (OP-GaP) crystal [48]. As both pulse trains were derived from a common oscillator, the generated MIR comb was offset-frequency free. The total emitted MIR power in the $7.25 - 8$ µm range was 0.8 to 2 mW. The comb repetition rate, $f_{rep}$, of nominally 125 MHz was tunable via a fiber stretcher placed in the oscillator and it was phase-locked to a reference signal synthesized by a radio-frequency (RF) generator referenced to a GPS-disciplined rubidium frequency standard.

The MIR comb beam was collimated and guided to a Herriot-type multi-pass absorption cell (Thorlabs, HC10L/M-M02) with an absorption path length of 10.436(15) m.[1] The cell was connected to a vacuum pump and a gas supply system providing $CH_4$ and $N_2$ sample gas (both from Air Liquide Gas AB, natural isotopic abundances, $CH_4$ purity: ≥99.995%). The pressure in the absorption cell was measured using a transducer with a resolution of 0.01 mbar (Leybold, CERAVAC CTR 101 N 100, uncertainty: 0.12%). After the spectra reported here had been recorded, a second pressure transducer with a higher resolution of 0.0001 mbar was installed (Leybold, CERAVAC CTR 100 N 1, uncertainty: 0.2%) and used to improve the calibration of the first pressure transducer. This procedure revealed an offset of the pressure indicated by the first transducer, used during recording the spectra, of 0.013(5) mbar. All pressure values reported below are corrected by this offset. The temperature was measured using an electronic thermometer with 0.1 K resolution placed on the optical table at a distance of roughly 20 cm from the multi-pass cell. We conservatively estimate the uncertainty of the temperature measurement to 1 K, taking into account a possible temperature gradient between the cell and the thermometer, as well as temperature drift during the acquisition of spectra.

Behind the absorption cell, the MIR beam was re-collimated and coupled into a fast-scanning FTS [41, 49]. The two out-of-phase interferometer outputs were detected by a pair of thermoelectrically-cooled HgCdTe detectors in a balanced configuration. A pair of adjustable irises placed in front of the detectors was used to balance the incident optical power and limit it to about 15 µW (in order not to exceed the linear response range of the detectors), as well as to block off-axis beam components to increase the interferogram contrast. The differential signal of the detector pair was digitized and recorded on a PC with a custom LabVIEW program. An interferogram of a narrow-linewidth CW diode-laser at 1.56 µm, whose beam was propagating on a path parallel to the comb beam through the FTS, served as a calibration standard for the FTS optical path difference.

We recorded $CH_4$ lines spanning a spectral range of more than 120 cm$^{-1}$ using two different OP-GaP crystals with poling periods of 58 and 60 µm, covering 1280 to 1380 cm$^{-1}$ and 1250 to 1330 cm$^{-1}$, respectively. The intensities of the lines reported in this work extend over more than four orders of magnitude. To achieve a sufficient signal-to-noise ratio (SNR) for as many lines as possible, we recorded data in six separate measurement runs (see Table 1). For observing the strong lines of the $^{12}CH_4$ $\nu_4$ fundamental band, we used a gas mixture of 5% $CH_4$ diluted in $N_2$ at a total pressure of 0.04 mbar. To measure the weaker lines of this band as well as those of the two $^{12}CH_4$ hot bands and the $\nu_4$ fundamental band of $^{13}CH_4$, we used pure $CH_4$ at pressures from 0.22 to 0.49 mbar. For the two runs with diluted $CH_4$, we first filled the previously evacuated absorption cell with $CH_4$ to a pressure of 5 mbar, added $N_2$ up to a total pressure of 100 mbar, then pumped the cell down to a total pressure of 0.04 mbar and isolated it from the vacuum and gas-supply system by closing a valve. For the measurements with pure $CH_4$, we first filled the cell to a few mbar of $CH_4$, then evacuated it to the desired pressure and closed the valve.

---

[1] Personal communication with Thorlabs Sweden AB.

For each run, we began the absorption measurement by locking the comb $f_{rep}$ and acquiring a first set of 50 interferograms. To reduce the sampling point spacing in the spectrum, we then stepped the $f_{rep}$ by tuning the RF generator in increments of 45 Hz, corresponding to 14 MHz in the optical domain, sufficient to yield around 8 points per full-width-half-maximum of the absorption lines, and acquired 50 interferograms at each of the nine steps needed to cover the entire comb mode spacing of 125 MHz. We repeated this $f_{rep}$ scanning cycle eight times in alternating directions to acquire a total of 400 interferograms at each $f_{rep}$ setting, except for run (1), where we acquired only 300 interferograms in six scanning cycles. One interferogram was acquired in 3.6 s, resulting in a total acquisition time of 3.6 h (2.7 h) for 400 (300) averages at each $f_{rep}$ step. To normalize the transmission spectra, we acquired a total of 400 [300 for run (1)] reference interferograms before and after the actual $CH_4$ absorption measurement, with the cell evacuated and the comb $f_{rep}$ stabilized to the first of the nine steps of the corresponding absorption measurement.

In the higher-frequency part of the spectral range studied [measurement runs (2), (5) and (6)], water absorption is strong. To minimize the effect of the anomalous water dispersion on the calibration of the optical path difference (OPD) in the FTS, we purged the FTS enclosure with dry air for measurement runs (2) and (5), which reduced the relative humidity to $\leq 4\%$. For run (6), we purged the FTS with pure nitrogen in order to investigate a possible effect of anomalous dispersion due to trace amounts of methane in the purge gas. This, however, did not influence the OPD calibration, and hence we conclude that trace amounts of methane do not significantly affect it.

**Table 1.** Measurement conditions and number of analyzed lines for each measurement run. Columns: 'Measurement run', designation of the respective measurement run; 'OP-GaP poling period', poling period of the OP-GaP crystal used in the comb source when recording the respective spectrum; 'Spectral coverage', spectral range over which lines have been analyzed; '$CH_4$ concentration', methane concentration in the sample gas (diluted in $N_2$ if applicable); 'Total pressure', total gas pressure in the absorption cell; 'Temperature', temperature measured in the vicinity of the cell; 'FTS purge gas', gas for purging the FTS enclosure if purged; 'Number of lines analyzed', number of absorption lines for which line parameters from the respective isotopologue and band(s) have been retrieved.

| Measurement run | OP-GaP poling period [µm] | Spectral coverage [cm⁻¹] | $CH_4$ concentration | Total pressure [mbar] | Temperature [°C] | FTS purge gas | Number of lines analyzed | | |
|---|---|---|---|---|---|---|---|---|---|
| | | | | | | | $^{12}CH_4\ \nu_4$ | $^{12}CH_4$ hot bands | $^{13}CH_4\ \nu_4$ |
| (1) | 60 | 1253 – 1322 | 5% | 0.04 | 22 | – | 110 | – | – |
| (2) | 58 | 1293 – 1373 | 5% | 0.04 | 23 | Dry air | 120 | – | – |
| (3) | 60 | 1252 – 1324 | 100% | 0.23 | 22 | – | 46 | 153 | 66 |
| (4) | 60 | 1250 – 1329 | 100% | 0.49 | 21 | – | 48 | 234 | 58 |
| (5) | 58 | 1280 – 1379 | 100% | 0.22 | 22 | Dry air | 101 | 164 | 89 |
| (6) | 58 | 1284 – 1378 | 100% | 0.49 | 22 | $N_2$ | 103 | 240 | 76 |

## 3   Spectral treatment and interleaving

The recorded data were processed with custom MATLAB scripts. We averaged the absolute values of the Fourier transformed interferograms for each $f_{rep}$ step. At this stage, we matched the sampling point positions to the comb line frequencies using the method described in Refs [40, 50] and elaborated on below, by minimizing the instrumental line shape (ILS) distortions caused by the interferogram truncation. We then obtained the transmission spectra by dividing the individual sample spectra by the corresponding reference spectrum smoothened and interpolated to the sampling points of the corresponding $f_{rep}$ steps. The background features in the comb envelope and spurious water absorption from the ambient air were broad enough to be unaffected by this smoothening and interpolation. We converted the transmission spectra to absorption spectra by means of the Lambert-Beer law. To remove the baseline remaining after the normalization process, we then fitted a spectral model based on parameters from the HITRAN2020 database [37] to the spectra, together with a baseline consisting of a 5th order polynomial and a set of sine terms added to model etalon fringes. We subsequently subtracted this baseline. Finally, we interleaved the spectra measured at the nine $f_{rep}$ steps to obtain a spectrum with a point spacing of ~14 MHz. Since significant discrepancies with respect to HITRAN were observed for some line center positions during the line fitting process (described in Section 4), the background subtraction and spectral interleaving were repeated with the spectral model updated with line positions from an initial line-by-line fit. This reduced the likelihood that mismatches between the model and the data distort the baseline near the lines shifted relative to HITRAN. Figure 1(a) shows the interleaved spectrum of the $^{12}CH_4$ $\nu_4$ band with data for the ranges 1250 – 1320 cm$^{-1}$ and 1320 – 1380 cm$^{-1}$ taken from measurement runs (1) and (2), respectively. The spectrum of the weaker bands, composed of data from measurement runs (5) and (6), with the same range limits, is shown in Figure 1(b). The maximum SNR in both measurements is about 880.

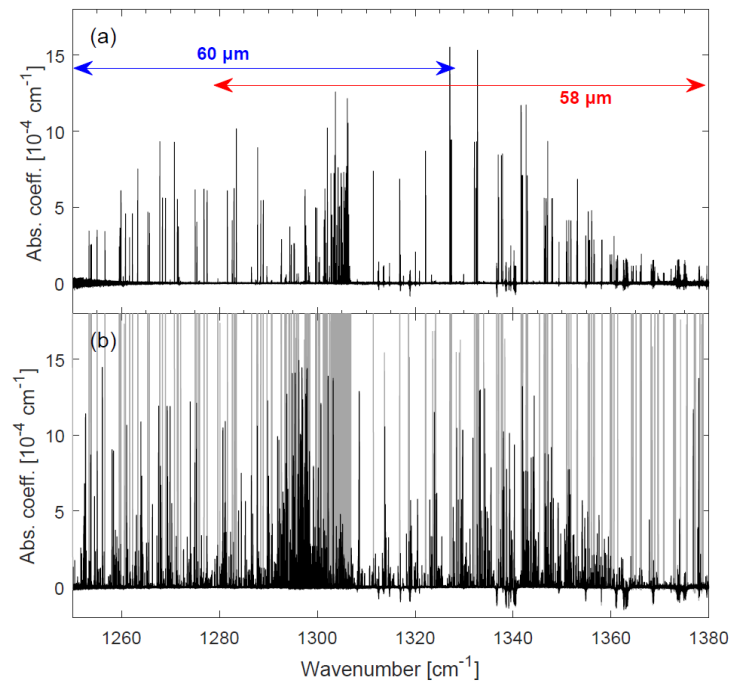

**Figure 1.** Interleaved methane spectra. (a) Spectrum of 5% CH$_4$ in N$_2$ at a total pressure of 0.04 mbar. The arrows indicate the spectral coverage of the comb with the two OP-GaP crystals with poling periods of 60 μm [measurement run (1)] and 58 μm [measurement run (2)]. Each data point shows the average of 300 or 400 individual spectra from run (1) or run (2), respectively, stitched together at 1320 cm$^{-1}$. (b) Spectrum of pure CH$_4$ at 0.49 mbar obtained from runs (4) and (6), 400 averages per data point, stitched together at 1320 cm$^{-1}$. The analyzed lines are shown in black, while the grey lines depict the stronger absorption lines shown in the upper panael, which are saturated at these conditions. (The distortions in the baseline showing negative excursions are due to residual water absorption in the ambient air.)

As explained in Refs [40, 50], the effective wavelength of the CW reference laser, $\lambda_{ref}$, used for OPD calibration is not known a priori with sufficient accuracy to match the sampling points to the comb mode frequencies because it depends on the alignment and collimation of the comb beam relative to the CW laser beam. Therefore, the matching is achieved by adjusting the value of $\lambda_{ref}$ in the analysis incrementally to minimize the ILS. To this end, we fitted Voigt line shapes to a subset of strong isolated absorption lines spread over the whole spectral range of each measurement using a range of $\lambda_{ref}$ values (in fractional increments of $10^{-8}$). For each line, the root-mean-square (rms) value of the fit residuals was plotted as a function of $\lambda_{ref}$ step and interpolated between the discrete increments to find the minimum, which corresponded to the optimum $\lambda_{ref}$ value (see Figure A.1 in Appendix A). We observed a clear dependence of the optimum reference wavelength on the line position in the spectrum, and concluded that the comb modes are mapped into the FTS spectrum with a non-uniform spacing. We attribute this wavelength-dependent spacing mainly to the divergence of the MIR beam since we observed a substantial change in this dependence after changing the OP-GaP crystal, which in turn changed the beam collimation after the crystal.

To avoid using different $\lambda_{ref}$ values for different positions in the spectrum, we followed a procedure described in more detail in Appendix A. Briefly, we identified ranges over which the wavenumber dependence of $\lambda_{ref}$ can be approximated as linear, and we applied a correction to $\lambda_{ref}$ to remove this linear dependence, followed by a global shift to the sampling points to match them to the comb mode positions. For the measurement runs (1), (3) and (4), this correction could be performed at once for the entire measured spectral range, while for the spectra from runs (2), (5) and (6) the local comb mode spacing varied significantly over the covered spectral region and $\lambda_{ref}$ could not be corrected in one step. Therefore, we divided these spectra into three [runs (2) and (5)] or four [run (6)] linear segments that we then corrected individually.

## 4    Line-by-line fitting

To determine the $CH_4$ line positions and intensities, we fitted Voigt line shapes to the absorption lines in the interleaved spectra, treating the line center positions, line intensities and Lorentzian widths as fit parameters, while fixing the Doppler widths to the calculated values around 115 MHz full width at half maximum (FWHM). The HITRAN2020 database [37] was used for line selection and assignment, as well as for providing the initial fit parameters. Lines listed in HITRAN within the spectral range covered by the light source were selected for fitting if they had a minimum line intensity corresponding to an SNR of ~20 in the measurement and were separated by at least 230 MHz from the nearest line also fulfilling this SNR criterion. The fitting window was chosen as ±460 MHz, or about 4 times the Doppler FWHM, around the HITRAN line positions. Lines separated by less than 1.5×460 MHz were fitted in a single window. In addition, lines with an SNR below 20 but above the noise floor were included in the fit model if they were within 460 MHz of the selected line, but their parameters were fixed to the HITRAN values. The measurements of the weak $^{12}CH_4$ hot bands and the $^{13}CH_4$ $v_4$ fundamental band also contained the strongly saturated lines due to (mainly) the $^{12}CH_4$ $v_4$ band. During line fitting, lines exceeding a peak absorption threshold of $1.5 \times 10^{-3}$ cm$^{-1}$ were therefore not fitted and were masked before fitting the weaker lines. We excluded lines from fitting if they were separated from masked lines by less than 740 – 850 MHz, depending on the respective measurement run.

The quality factor, defined as the peak absorption value divided by the standard deviation of the residuals, was calculated for each fitted line. The standard deviation of the residuals was evaluated within the single line fitting window, i.e., ±460 MHz. We subsequently rejected lines with a quality factor lower than 20. Figure 2 shows an example of a fit to a line of the $^{12}CH_4$ $v_4$ band, with a quality factor of 260.

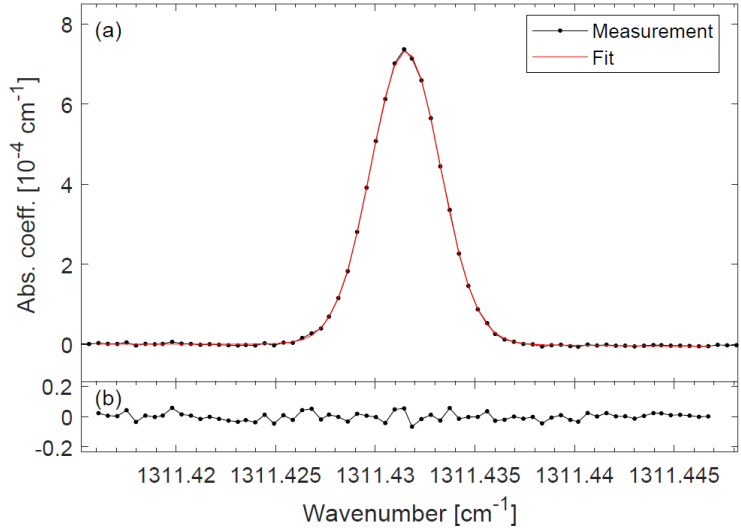

**Figure 2.** (a) Absorption spectrum of the $(J,C,\alpha)$: $(1, A_2, 1) \leftarrow (0, A_1, 1)$ line of the $^{12}CH_4$ $\nu_4$ band at a concentration of 5% and a total pressure of 0.04 mbar (black markers) together with a fitted Voigt line profile (red line). (b) Fit residuals.

The linewidth is dominated by the Doppler broadening at the gas pressures of all measurement runs. The pressure-broadening contribution, calculated based on the HITRAN parameters, varies between ~0.2 MHz FWHM at 0.04 mbar and ~2.1 MHz FWHM at 0.49 mbar. However, the average and the standard deviation of the Lorentzian widths of all fitting lines from all the measurement runs were 2.5 MHz and 2.9 MHz, respectively. We attribute this wide spread of the homogeneous widths to the fact that the SNR of most of the lines is of the order of the ratio of the Doppler to the Lorentzian widths, which limits the precision with which the latter can be determined. Moreover, we previously observed excess line broadening [41], whose origin remains unclear. Possible causes include the spectral widths of the MIR comb modes and distortions of the wavefront of the MIR beam, caused, e.g., by the multi-pass absorption cell. We currently have no means to measure the comb mode width and cannot quantify the influence of the wavefront shape on the measured line widths. However, as there is no asymmetry in the line shapes (the fit residuals are flat), and given the good agreement of our previously determined line positions of the $N_2O$ $\nu_1$ band [41] with those independently obtained by AlSaif et al. [42] from comb-referenced CW laser spectroscopy, we conclude that the value of pressure broadening does not affect the retrieved line positions beyond the estimated uncertainty limits (see Section 5.1). For intensities, however, we take into account the possible influence of the homogenous widths by flagging the lines with largely deviating widths in the line list, as described in Sections 5.2 and 6.

To obtain the line intensities at the HITRAN reference temperature (296 K), we converted the values retrieved from line fitting to 296 K by means of Eq. (9) from Ref. [51] using the temperatures given in Table 1 together with the lower-level energies from Ref. [37] and the internal partition sum from Ref. [52].

## 5 Uncertainty estimates and averaging of line parameters

### 5.1 Line positions

We identify two sources of uncertainty in the retrieved line positions. The first one is the uncertainty in the fit of the Voigt line profiles, which is specific to each line and depends mostly on its SNR. This uncertainty is in the range of 0.06 MHz to 2.2 MHz for the lines from the $^{12}CH_4$ $\nu_4$ and $2\nu_4 - \nu_4$ bands, and the $^{13}CH_4$ $\nu_4$ band, and 0.3 MHz to 1.8 MHz for the $^{12}CH_4$ $\nu_2 + \nu_4 - \nu_2$ band.

The second contribution stems from the uncertainty in the determination of the reference wavelength $\lambda_{ref}$, described in Section 3 and in Appendix A, and is common for all lines in a given measurement or measurement segment. We estimated this contribution based on the spread of the $\lambda_{ref}$ values found for the subset of lines that were used to optimize $\lambda_{ref}$ for the respective measurement or measurement segment. Namely, we calculated the range (maximum – minimum) of the optimum $\lambda_{ref}$ found for these lines (in terms of fractional increments of $10^{-8}$) and converted it to a one-sided $1\sigma$ uncertainty by multiplying it with $\text{erf}(1/\sqrt{2})/2 \approx 0.34$ (with erf the Gauss error function). We then rounded this value up to the nearest half-integer to take into account the discrete nature of the increments and converted it to the corresponding shift in line center positions which was ~0.2 MHz per $\lambda_{ref}$ step. This contribution to the total line center uncertainties was between 0.2 MHz and 0.6 MHz, depending on the measurement or measurement segment.

The two uncertainty contributions were summed in quadrature for each retrieved line position in each of the measurement runs. More than half of the analyzed lines were measured in more than one measurement run. For these lines, we averaged the vaues obtained from the repeated measurements weighted by the inverse of the squared uncertainty of each value (obtained as described above), and estimated the uncertainty of the resulting weighted mean by propagating the uncertainties of the individual values.

An additional contribution to the uncertainty of the position of certain lines arises from line profile distortions caused by nearby interfering lines. As mentioned in Section 0, lines too weak to be fitted, but located within the fitting window of a stronger fitted line, were included in the fit model with their line parameters fixed to the corresponding HITRAN2020 values. Since these values might be inaccurate, the fit to the stronger line might be inaccurate as well. The lines whose positions could be affected by this hard-to-quantify effect, i.e. lines that have at least one interfering line within a range of $\pm 230$ MHz (about two Doppler FWHM) with an intensity of at least 0.1% of the listed line, are flagged in the line lists (see Section 6).

The reported line positions are not corrected for pressure shifts. For the two measurements with diluted $CH_4$ [measurement runs (1) and (2)] the absolute pressure shifts are expected to be <0.02 MHz based on the shift coefficients for $CH_4$ in $N_2$ reported by Rinsland et al. [53]. Since the minimum reported uncertainty from these measurement runs is 0.21 MHz, these pressure shifts are negligible. For the measurements with pure $CH_4$, the absolute shifts of the $^{12}CH_4$ ($^{13}CH_4$) $\nu_4$ lines are expected to be <0.09 MHz (<0.06 MHz) for runs (3) and (5), and <0.2 MHz (<0.1 MHz) for runs (4) and (6), based on the $CH_4$ self-shift coefficients reported by Smith et al. [54, 55]. With minimum estimated uncertainties of 0.19 MHz for the $^{12}CH_4$ and $^{13}CH_4$ $\nu_4$ lines retrieved from these measurement runs, the pressure shifts cannot a priori be neglected. They could, e.g., be modeled with the 'off-diagonal relaxation matrix elements' formalism described by Smith et al. [22], which, however, is beyond the scope of this work. For the two $^{12}CH_4$ hot bands, we are not aware of any published self-pressure-shift coefficients and hence cannot estimate the pressure shifts.

## 5.2    Line intensities

For  measurement runs (3) to (6) with pure $CH_4$, the main sources of uncertainties in the line intensities are the line-profile fitting, and the measurements of sample pressure and temperature. The uncertainty of the absorption path length (<0.2%) is negligible. The relative uncertainties of the fitted line intensities are in the range of 0.15% to 6.6% (on average 1.7%). The uncertainty of the sample gas pressure includes contributions from the finite resolution of the pressure gauge and from the offset calibration. The former results in a relative uncertainty of 2% for run (4) and (6), and 4% for run (3) and (5), while the latter accounts to 1% and 2%, respectively. The uncertainty contribution from the gas temperature varies for each spectral line as it depends on the respective lower-level energy. While essentially vanishing for low-$J$ fundamental-band transitions, it rises to about 2-3% for high-$J$ fundamental and hot-band lines, for a temperature uncertainty of $\pm 1$ K. All uncertainty contributions are summed in quadrature. Similarly as for the line positions, we obtained the final line intensities of lines measured multiple times as the mean of the individual values weighted by the inverse of their squared uncertainties, and the final uncertainties by propagating the uncertainties of the individual values accordingly.

For measurement runs (1) and (2) with diluted $CH_4$, an additional source of uncertainty in the line intensities arises from the imperfect mixing of $CH_4$ and $N_2$ in the gas system. Since we cannot quantify the resulting uncertainty in the $CH_4$ concentration, we do not report line intensities from these runs.

As mentioned in Section 4, the spread of the fitted homogenous widths was unphysically large. The influence of this spread on the uncertainty of the retrieved intensities is difficult to quantify. Therefore we flag in the line list the lines for which the fitted homogeneous width of at least one measurement deviates by more than three times the fit uncertainty from the mean value of the corresponding measurement run.

# 6    Experimental line lists

The experimental line lists for each band are available as Supplementary Material. For each line, the line-center position, the uncertainty of the line position and the upper- and lower-state quantum labels (following the HITRAN notation) are given. For lines for which line intensities have been determined, the line intensity is listed together with the corresponding relative uncertainty. The lines whose fitted homogeneous width deviates by more than three times the retrieved fit uncertainty from the mean of the corresponding measurement run are flagged with stars. Furthermore, for lines with close-by interfering lines the intensity ratio of the interfering versus the listed line is given as a code: 10 for >10%; 5 for 5% to 10%; 2 for 2% to 5%; 1 for 1% to 2%; 0 for 0.1% to 1%.

Table 2 summarizes the key figures of the line lists for the four analyzed bands, i.e., the covered spectral range, the total number of fitted lines, and the number of lines measured multiple times. Also listed are the median uncertainties, as well as the uncertainty ranges, of the line positions and intensities. We note that from the total of 322 analyzed $^{12}CH_4$ $v_4$ lines, we report line positions and intensities of 157 lines, while for the remaining 165 lines, we report only line positions because of the above-mentioned imperfect mixing of the diluted gas samples.

**Table 2**: Key figures of the experimental line lists for the bands and isotopologues studied. Columns: 'Isotopologue, Band', designation of the respective $CH_4$ isotopologue and the vibrational band; 'Spectral range analyzed', spectral range over which lines of the respective isotopologue and band have been analyzed; 'Lines analyzed', number of lines analyzed from the respective band and isotopologue; 'Lines measured repeatedly', number of lines that have been measured repeatedly (twice, three or four times); 'Line position uncertainties', median values and range of the uncertainties in the measured line positions; 'Relative line intensity uncertainties', median values and range of the relative uncertainties in the measured line intensities.

| Isotopologue, Band | Spectral range analyzed [cm⁻¹] | Lines analyzed | Lines measured repeatedly | | | Line position uncertainties [MHz] | | Relative line intensity uncertainties [%] | |
|---|---|---|---|---|---|---|---|---|---|
| | | | 2x | 3x | 4x | Median | Range | Median | Range |
| $^{12}CH_4$ $v_4$ | 1252 – 1379 | 322(*) | 154 | 5 | 14 | 0.36 | 0.19 – 2.3 | 2.5 | 1.6 – 7.7 |
| $^{12}CH_4$ $2v_4 - v_4$ | 1250 – 1365 | 313 | 141 | 59 | 35 | 0.47 | 0.20 – 2.0 | 2.9 | 1.8 – 6.8 |
| $^{12}CH_4$ $v_2 + v_4 - v_2$ | 1262 – 1361 | 77 | 21 | 8 | – | 0.98 | 0.45 – 1.8 | 4.1 | 2.3 – 6.7 |
| $^{13}CH_4$ $v_4$ | 1252 – 1367 | 131 | 66 | 10 | 24 | 0.29 | 0.19 – 1.8 | 2.1 | 1.5 – 7.3 |

(*) For the $^{12}CH_4$ $v_4$ band, line intensities have been retrieved only for the 157 lines observed in the measurement runs (3) to (6).

## 7 Global model

### 7.1 Line positions

The high-precision line parameters determined in this work can be used to improve the global modeling of the methane absorption spectrum. We thus included the new line positions in the methane global fit using the tensorial model developed in the Dijon group described in Ref. [25]. This model groups the methane vibrational levels into polyads [36] denoted $P_0$, $P_1$, $P_2$ … where $P_0$ is the ground vibrational state, $P_1$ is the $v_2/v_4$ dyad, $P_2$ is the $v_1/v_3/2v_2/2v_4/v_2+v_4$ pentad, etc. All line positions retrieved in this work have been included in the model; line positions that were already present in the fit were replaced and new line positions were added. For the $P_1$-$P_0$ dyad transitions of $^{12}CH_4$, this amounts to 322 $v_4$ lines (Table 2), including 40 new lines not present in Ref. [25]. For the $P_2$-$P_1$ hot bands of $^{12}CH_4$, we included 390 lines (313 $2v_4 - v_4$ lines and 77 $v_2 + v_4 - v_2$ lines, see Table 2). For the $P_1$-$P_0$ dyad transitions of $^{13}CH_4$, 3 of the 131 $v_4$ lines (Table 2) were not present in the most recent global fit of the $^{13}CH_4$ spectrum [23] while the remaining ones replaced previous measurements. The residuals of the global fits including the new data are shown in Figure 3 and Figure 4 for the $P_1$-$P_0$ dyad and $P_2$-$P_1$ hot bands of $^{12}CH_4$, respectively, and in Figure 5 for the $P_1$-$P_0$ dyad of $^{13}CH_4$. In those figures, the new data are shown by red crosses, while the previous absorption and emission data from Refs. [23, 25] (and references therein) are shown by blue circles and green diamonds, respectively. The uncertainties of the new data are visibly reduced compared to the previous absorption and emission data. The previous absorption data that has been replaced by the new measurements have uncertainties ranging from 6 to 90 MHz, and their medians are 6 MHz for the $P_1$-$P_0$ dyad, and 30 for the $P_2$-$P_1$ dyad, respectively, compared to the median uncertainty of 0.33 MHz and 0.51 MHz of the new $P_1$-$P_0$ and $P_2$-$P_1$ dyad lines, respectively.

The global rms deviation for $P_1$-$P_0$ transitions (see Figure 3) is $2.72 \times 10^{-3}$ cm$^{-1}$ / 81.4 MHz, only slightly smaller than in Ref. [25] ($2.83 \times 10^{-3}$ cm$^{-1}$ / 84.9 MHz), but the residuals for lines from this work are greatly reduced: the rms deviation of the new lines is only $0.065 \times 10^{-3}$ cm$^{-1}$ / 1.94 MHz, 2.7 times smaller than for the remaining (i.e. not replaced by the new data) $P_1$-$P_0$ absorption lines ($0.178 \times 10^{-3}$ cm$^{-1}$ / 5.34 MHz) and 69 times smaller than for the remaining $P_1$-$P_0$ emission lines ($4.46 \times 10^{-3}$ cm$^{-1}$ / 134 MHz). The rms deviation of the new lines that replaced previous absorption and emission data is $0.029 \times 10^{-3}$ cm$^{-1}$ / 0.871 MHz and $0.165 \times 10^{-3}$ cm$^{-1}$ / 4.95 MHz, respectively, reduced by a factor of 8.2 and 21 compared to the replaced absorption ($0.239 \times 10^{-3}$ cm$^{-1}$ / 7.17 MHz) and emission ($3.40 \times 10^{-3}$ cm$^{-1}$ / 102 MHz) data, respectively.

The improvements are similar for the $P_2$-$P_1$ hot bands (see Figure 4). The global rms deviation is reduced only slightly from $4.52 \times 10^{-3}$ cm$^{-1}$ / 136 MHz to $4.41 \times 10^{-3}$ cm$^{-1}$ / 132 MHz, but the residuals of the new lines ($0.114 \times 10^{-3}$ cm$^{-1}$ / 3.41 MHz) are improved by a factor of 8 compared to the remaining previous absorption data ($0.913 \times 10^{-3}$ cm$^{-1}$ / 27.4 MHz) and a factor of 44 compared to the remaining emission data ($5.03 \times 10^{-3}$ cm$^{-1}$ / 151 MHz). Compared to the absorption and emission data that have been replaced, the residuals of the new lines are improved by a factor of 8 and 19, respectively ($0.718 \times 10^{-3}$ cm$^{-1}$ / 21.5 MHz and $2.62 \times 10^{-3}$ cm$^{-1}$ / 78.5 MHz for the previous absorption and emission measurements, respectively, versus $0.090 \times 10^{-3}$ cm$^{-1}$ / 2.70 MHz and $0.141 \times 10^{-3}$ cm$^{-1}$ / 4.24 MHz for the corresponding sets of lines in the revised dataset).

In the case of $^{13}CH_4$ (see Figure 5), there is a factor of more than 3 improvement between remaining previous ($0.070 \times 10^{-3}$ cm$^{-1}$ / 2.1 MHz) and present ($0.021 \times 10^{-3}$ cm$^{-1}$ / 0.63 MHz) absorption lines.

The uncertainty on some of the effective Hamiltonian parameters is also slightly improved. The new effective Hamiltonian parameter list is included as Supplementary Material to the present paper, in the same way as it was done for Ref. [25].

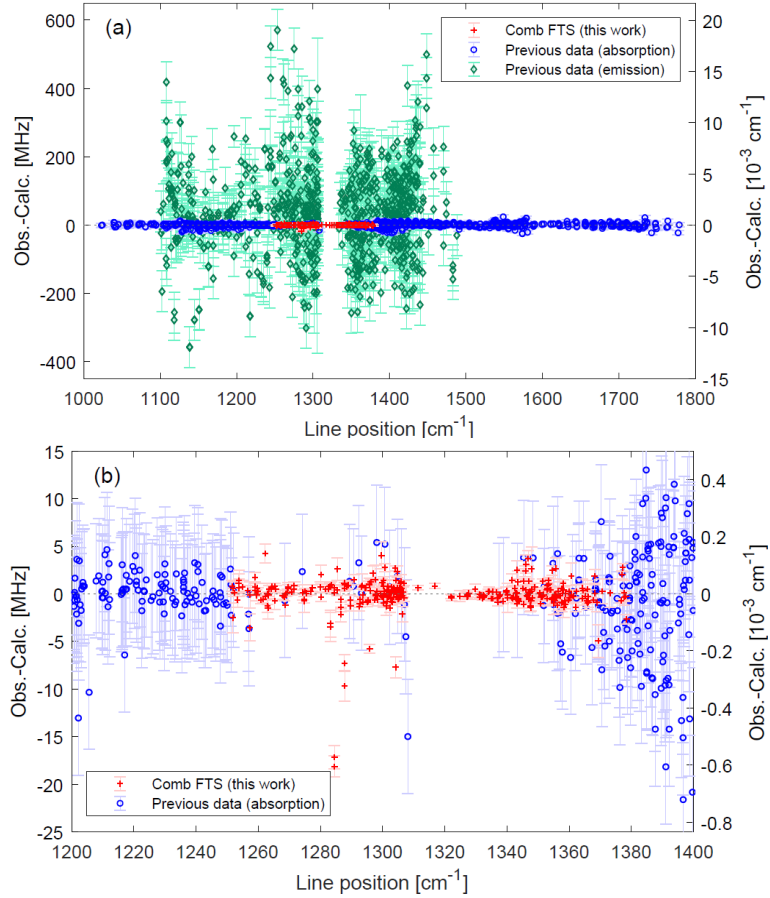

**Figure 3.** (a) The residuals of the global fit for the $^{12}CH_4$ $P_1$-$P_0$ line positions obtained from optical frequency comb FTS absorption measurements reported in this work ($v_4$ band, red crosses) as well as from previous absorption (blue circles) and emission (green diamonds) spectroscopy studies [25]. (b) Detailed view of the spectral range including the lines reported in this work. The reduced uncertainty of the comb FTS measurements as compared to previous absorption measurements is evident. (For clarity, previous emission measurements are not shown.)

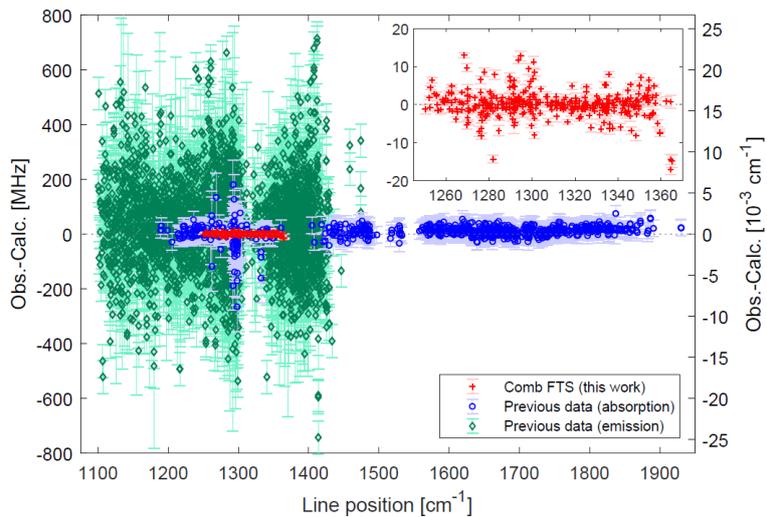

**Figure 4.** The residuals of the global fit for the $^{12}CH_4$ $P_2$-$P_1$ line positions from optical frequency comb FTS absorption measurements reported in this work ($2v_4 - v_4$ and $v_2 + v_4 - v_2$ bands, red crosses) as well as from previous absorption (blue circles) and emission (green diamonds) spectroscopy studies [25]. Inset: detailed view of the spectral lines reported in this work.

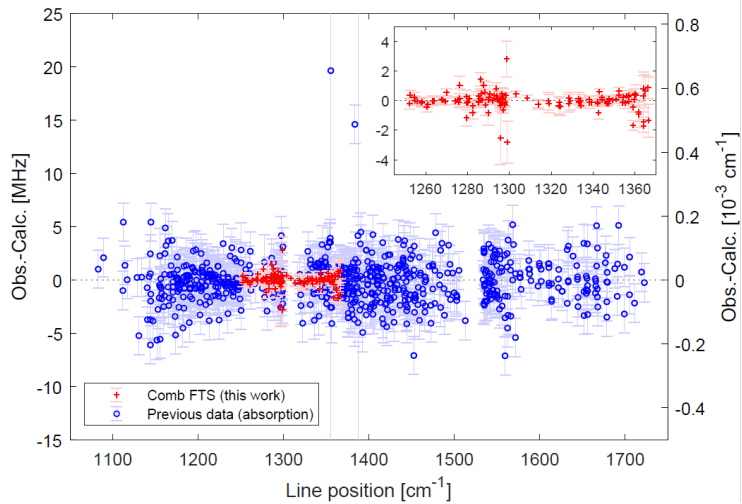

**Figure 5.** The residuals of the global fit for the $^{13}CH_4$ $P_1$-$P_0$ line positions from optical frequency comb FTS absorption measurements reported in this work ($\nu_4$ band, red crosses) as well as from previous absorption measurements (blue circles) [23]. Inset: detailed view of the spectral lines reported in this work

## 7.2 Line intensities

We also included line intensities from this work in the effective dipole moment parameter fits (as in Ref. [25]) for the $P_1$-$P_0$ dyad and $P_2$-$P_1$ hot bands of $^{12}CH_4$. Since the precision of line intensities in this work is similar to that of one of previous methane studies, we only included lines that were not included in the previous intensity fit. This amounts to 62 new $\nu_4$ intensities, 33 new $\nu_2 + \nu_4 - \nu_2$ intensities and 140 new $2\nu_4 - \nu_4$ intensities. The results are shown in Figure 6 and Figure 7 for $P_1$-$P_0$ and $P_2$-$P_1$ lines, respectively. The global rms deviations are very similar to the previous ones (3.3% in both cases for the $P_1$-$P_0$ dyad and 4.9% compared to 4.6% for the $P_2$-$P_1$ dyad). The advantage of this inclusion, however, is to increase the data set for intensity fits with several tens of new lines.

These results will be used to update the MeCaSDa database of calculated methane lines (http://vamdc.icb.cnrs.fr) described in Ref. [56].

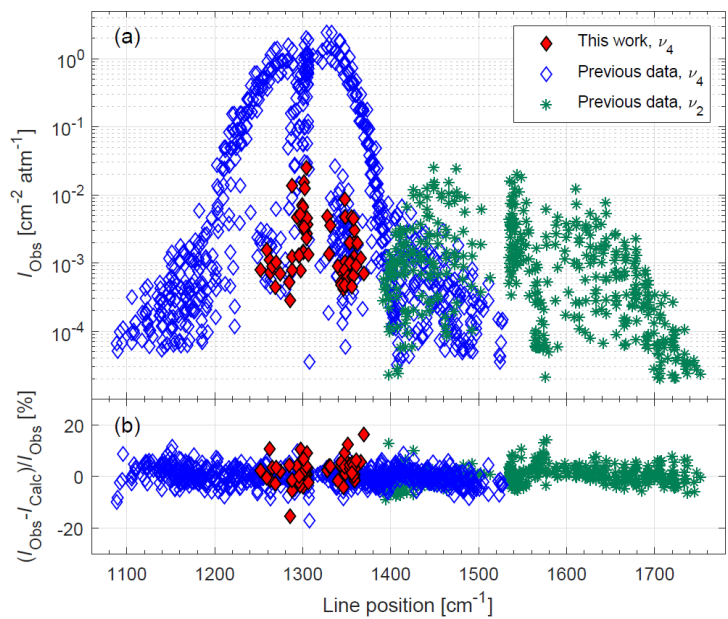

**Figure 6.** (a) Observed line intensities and (b) relative fit residuals of the $^{12}CH_4$ $P_1$-$P_0$ transitions obtained using the same global fit procedure as in Ref. [25]. Red, filled diamonds show transitions from the present work ($\nu_4$ band) included in the global fit of the effective dipole operators, while open blue diamonds and green stars show transitions from previous measurements of the $\nu_4$ and $\nu_2$ band, respectively [25].

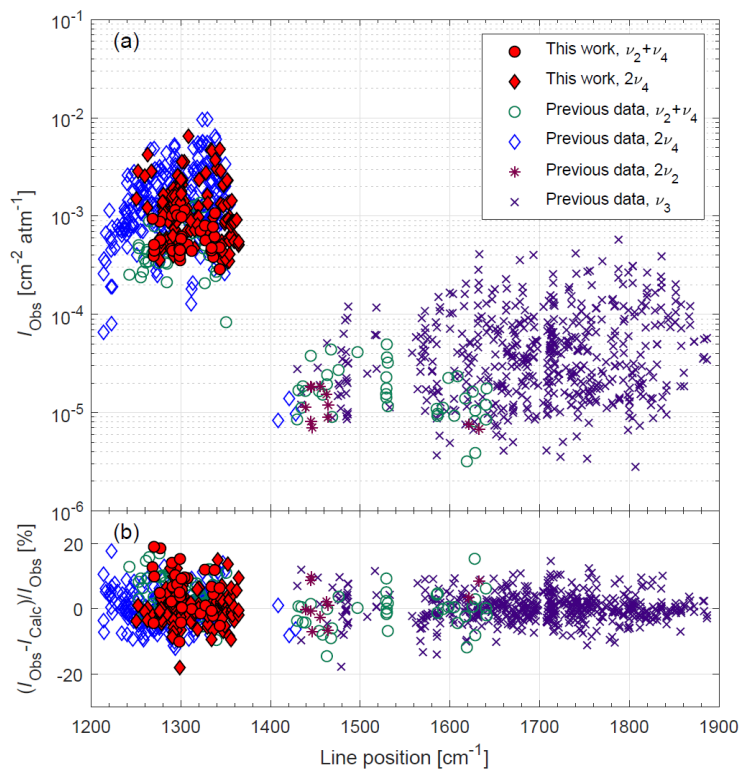

**Figure 7.** (a) Observed line intensities and (b) relative fit residuals of the $^{12}CH_4$ $P_2$-$P_1$ transitions using the same global fit procedure as in Ref. [25]. Red filled symbols show transitions from the present work included in the global fit of the effective dipole operators, while open symbols, stars and crosses show transitions from previous measurements [25], as marked in the legend. Transitions are labeled by their respective upper vibrational level.

## 8    Summary and Conclusions

In this work, we report a new experimental line list of the $\nu_4$ fundamental bands of $^{12}CH_4$ and $^{13}CH_4$ and two hot bands of $^{12}CH_4$ obtained from line-by-line fitting to high-resolution spectra recorded using a Fourier transform spectrometer based on a MIR optical frequency comb. The uncertainties of line positions are in the range of 0.19 to 2.3 MHz, significantly lower than those from previous absorption data. Incorporating all 843 new line positions in the global model of the methane spectrum adds 43 lines to the previous dataset [23, 25] and results in a reduction of the fit rms residuals by a factor of 8 and 20 compared to the replaced previous absorption and emission data. The line position uncertainty is limited mostly by the SNR in the spectra, and can thus be improved by longer averaging. The line intensities are reported for 678 lines (of which 235 were previously not included in the global methane model) with uncertainties similar to those previously available. These uncertainties can be reduced by increasing the SNR in the spectra, and a more accurate measurement of sample pressure and temperature. The highly-accurate line positions obtained from comb-based Fourier transform spectroscopy allow significant improvements to the global spectral model, and further improvements are expected when the technique is applied to measure spectra in other parts of the methane spectrum.

## Acknowledgments

The authors thank Iouli Gordon for pointing out the need for more accurate $CH_4$ line position measurements in the 8 μm spectral range, and Thorlabs Sweden for the loan of the multi-pass cell.

## Funding

This work was supported by the Knut and Alice Wallenberg Foundation (KAW 2015.0159 and 2020.0303), the Swedish Research Council (2016-03593 and 2020-00238), and the Foundation for Polish Science (First TEAM/2017-4/39).

## Appendix A – FTS frequency scale optimization

The frequency of the $n^{th}$ mode of a carrier-envelope-offset free optical frequency comb is given by

$$\nu_{OFC} = n f_{rep} \, . \tag{A.1}$$

The frequency scale of the spectrum resulting from taking the discrete Fourier transform of an interferogram in turn has the general form

$$\nu_{FTS} = n f_{FTS} \, , \tag{A.2}$$

with a sampling point spacing

$$f_{FTS} = \frac{qc}{2\lambda_{ref} N_0} \, , \tag{A.3}$$

where c is the speed of light in vacuum, $\lambda_{ref}$ the effective wavelength of the CW reference laser, q the number of points sampled per $\lambda_{ref}$ in the OPD domain (here q=4 ) and $2N_0$ the number of data points in the interferogram. The first step in the analysis is to match $f_{FTS}$ to $f_{rep}$ by adjusting the length of the interferogram. This cannot be done with arbitrary precision since $N_0$ is restricted to integer values. Thus $N_0$ in general deviates from the (non-integer) value yielding ideal matching N* and is related to it by

$$N^* = N_0 (1 + \varepsilon_0) \, , \tag{A.4}$$

where the deviation is contained in the parameter $\varepsilon_0$ . Furthermore, the effective reference wavelength is unknown and the approximate value used in the analysis is denoted by

$$\lambda'_{ref} = \lambda_{ref} (1 + \eta) \, , \tag{A.5}$$

where the deviation from the true value is expressed by the unknown parameter $\eta$ . For the reference wavelength $\lambda'_{ref}$ , $f_{rep}$ can be expressed as

$$f_{rep} = \frac{qc}{2\lambda'_{ref} N^*} = \frac{qc}{2\lambda_{ref}(1+\eta)N_0(1+\varepsilon_0)} \, . \tag{A.6}$$

Solving Eq. (A.6) for $N_0$ and substituting that value into Eq. (A.3), yields the corresponding point spacing

$$f_{FTS} = f_{rep}(1+\varepsilon_0)(1+\eta) = f_{FTS}^0 (1+\eta) \, , \tag{A.7}$$

where $f_{FTS}^0 = f_{rep}(1+\varepsilon_0)$ . The point spacing $f_{FTS}$ deviates from $f_{rep}$ due to the integer value of $N_0$ and uncertainty in $\lambda'_{ref}$ , and this mismatch leads to an offset between sampling points and comb modes that increases with the index n. The parameter $\varepsilon_0$ is known for a given $\lambda'_{ref}$ from Eq. (A.6), and the offset due to it can be corrected at a given index $n_{opt}$ by introducing a shift to the FTS scale

$$\nu_{FTS} = n f_{FTS}^0 + f_{shift} \, , \tag{A.8}$$

where $f_{shift} = -n_{opt}\varepsilon_0 f_{rep}$ (see Ref. [50]). The effect of integer $N_0$ can hence be minimized for modes in the vicinity of the mode $n_{opt}$. In the final interleaved spectrum, $N_0$ is different for the constituent spectra measured at the different $f_{rep}$ steps, and $\varepsilon_0$ accordingly varies between its maximum and minimum value of $\pm 1/(2N_0)$ (see Ref. [50]). The small value of this parameter and the fact that it is different for consecutive sampling points on an absorption line profile implies that it does not cause a shift in the line positions in the final interleaved spectrum. It is hence neglected from here on and the FTS sampling point spacing is written as

$$f_{FTS} = f_{rep}(1+\eta) . \tag{A.9}$$

To find the optimum value of $f_{FTS}$ (minimum $\eta$), $\lambda'_{ref}$ is varied to minimize the ILS distortions at a particular absorption line [50]. Figure A.1 shows a plot of the rms value of the fit residuals of one particular line used to determine the optimum $\lambda'_{ref}$ value, as a function of the reference wavelength expressed in relative incremental steps of $10^{-8}$. The curve shows an interpolation of the data points, and the green marker indicates the minimum value of the increment, corresponding to the optimum value of $\lambda'_{ref}$.

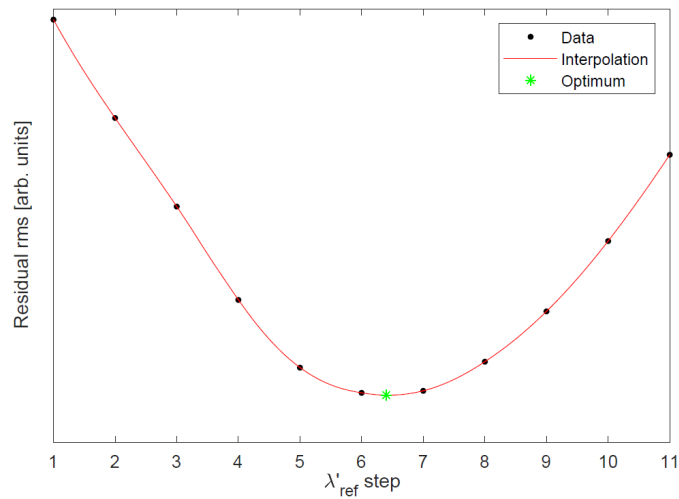

Figure A.1. The rms value of the line fit residuals (black dots) for different values of $\lambda'_{ref}$ for a single line where one step corresponds to a relative change of $\lambda'_{ref}$ of $10^{-8}$. The red curve is an interpolation of the discrete steps and its minimum value (green marker) corresponds to the optimum $\lambda'_{ref}$ value.

The procedure described above works well for individual absorption lines. However, $\lambda'_{ref}$ depends on the relative alignment of the comb and reference beam and their collimation, which can be different for different parts of the comb spectrum. Even though the comb mode frequencies are equidistant, they are mapped non-linearly to wavelengths in the FTS spectrum. This is schematically depicted in Figure A.2 where red dots show the mapped comb mode frequencies, $\nu'_{OFC}$. The (exaggerated) nonlinearity implies that it is not possible to match the sampling points to the comb modes in the entire measured spectral range by the optimization routine described so far, due to the local difference in slopes at the intersections between the FTS scale and the comb mode positions. This is illustrated by two (linear) FTS scales $\nu^A_{FTS}$ and $\nu^B_{FTS}$, that intersect the comb mode curve at indices $n_A$ and $n_B$, shown by black dots. Below we describe a correction procedure that allows calculating a local FTS scale, $\nu^{corr}_{FTS}$, with smallest deviation from the comb scale in the selected range from $n_A$ to $n_B$.

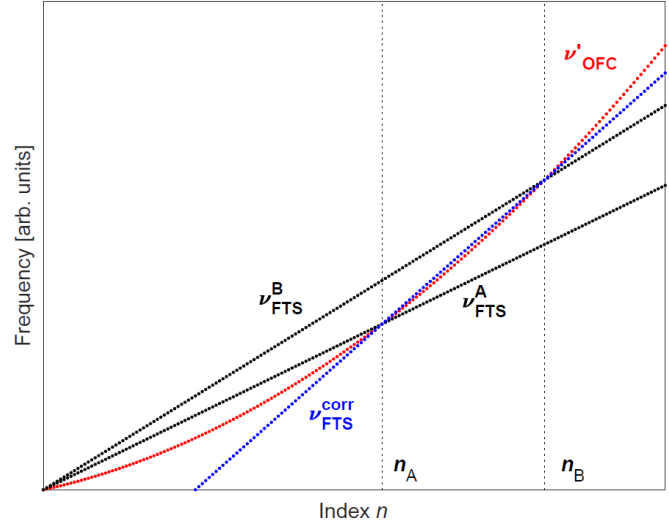

Figure A.2. Graphic depiction of the effect of nonlinear mapping of the optical comb frequencies in the FTS. The horizontal coordinate corresponds to the integer indices of the comb modes and the sampling points. The comb modes (red dots) occur along a curved line (curvature highly exaggerated here). The black dots indicate the FTS sampling points for two different values of the reference wavelength which translate into different point spacings reflected by the different slopes of these lines. The sampling points of the black curves coincide with the comb modes at indices $n_A$ and $n_B$. The comb modes frequencies between $n_A$ and $n_B$ are approximated with the corrected FTS scale (blue dots).

Suppose that a value of the reference laser wavelength $\lambda_{ref}^A$ has been found that minimizes the fit residual rms of an absorption line at a comb mode with index $n_A$, hence corresponding to an FTS scale $\nu_{FTS}^A = n f_{FTS}^A$. At comb mode index $n_B$, a different optimumu value $\lambda_{ref}^B = \lambda_{ref}^A(1+\Delta\eta)$ is found, corresponding to $\nu_{FTS}^B = n f_{FTS}^B$. The two sample point spacings will then be related to each other by $f_{FTS}^B = f_{FTS}^A(1+\Delta\eta)$. The point spacing can be matched to the average comb mode spacing between $n_A$ and $n_B$ through the relation

$$f_{FTS}^{corr} = \frac{f_{FTS}^B n_B - f_{FTS}^A n_A}{n_B - n_A} = \frac{f_{FTS}^A\left[(1+\Delta\eta)\nu_B - \nu_A\right]}{\nu_B - \nu_A} = f_{FTS}^A\left(1 + \frac{\nu_B\Delta\eta}{\nu_B - \nu_A}\right). \tag{A.10}$$

In the second step, the indices have been replaced by the corresponding optical frequencies $\nu_{A/B} = f_{rep} n_{A/B}$. This corrected point spacing corresponds to the reference wavelength

$$\lambda_{ref}^{corr} = \lambda_{ref}^A\left(1 + \frac{\nu_B\Delta\eta}{\nu_B - \nu_A}\right). \tag{A.11}$$

Using $f_{FTS}^{corr}$ will shift the sampling points away from the comb modes unless an additional frequency shifting correction $f_{corr}$ is introduced to the FTS scale according to

$$\nu_{FTS}^{corr} = n f_{FTS}^{corr} + f_{shift} + f_{corr}, \tag{A.12}$$

where the correction $f_{corr}$ is calculated in order to match the comb modes and sampling points at mode $n_A$ (and consequently at $n_B$ as well) as

$$f_{corr} = f_{FTS}^A n_A - f_{FTS}^{corr} n_A = f_{FTS}^A n_A\left[1 - \frac{(1+\Delta\eta)\nu_B - \nu_A}{\nu_B - \nu_A}\right] = -\nu_A\left[\frac{\nu_B\Delta\eta}{\nu_B - \nu_A}\right]. \tag{A.13}$$

The resulting corrected FTS scale is shown by the blue dots in Figure A.2. The choice of the frequencies $\nu_A$ and $\nu_B$ is determined by the range over which the curve $\nu'_{OFC}$ in Figure A.2 can be approximated with a linear slope. This depends on the local curvature of $\nu'_{OFC}$ relative to the uncertainty in determining the optimum $\lambda'_{ref}$ (see below for practical procedure). Ideally the curvature is small enough that a whole measurement can be corrected at once and $\nu_A$ and $\nu_B$ are accordingly chosen close to the edges of the measured range. However, if the curvature is too large, it is necessary to divide the measurement into segments and correct them individually.

Figure A.3(a) shows the results of a line-by-line optimization of $\lambda'_{ref}$ for measurement run (4) [see Table 1] before the scale correction process, with the vertical axis expressed in terms of the $\lambda'_{ref}$ step (relative increment $10^{-8}$). This serves as the basis for the correction process. A linear fit (red line) to the data points is used to determine $\Delta\eta$. Note that the line fits well within the scatter of the data and hence can be used for the entire measured range. We then calculate $\lambda^{corr}_{ref}$ and $f_{corr}$ from Eqs. (A.11) and (A.13). Redoing the analysis using $f_{corr}$ and scanning the reference wavelength around $\lambda^{corr}_{ref}$ yields the plot in Figure A.3(b) confirming that the linear slope has been cancelled. Thus, with the final reference wavelength taken as the mean of the data points (solid red line), the FTS scale in Eq. (A.12) matches the comb modes to within the achievable precision. The uncertainty contribution to the line positions stemming from this process is estimated from the $1\sigma$ uncertainty range in $\lambda^{corr}_{ref}$ (red dashed lines) as described in Section 5.1.

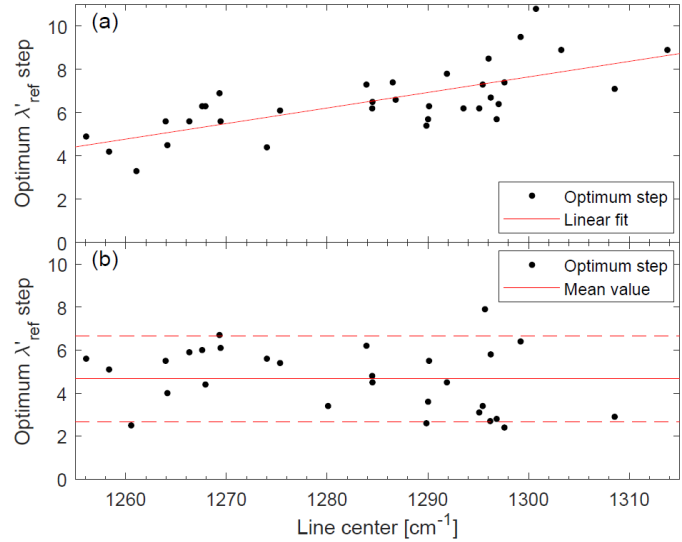

Figure A.3 (a) The optimum $\lambda'_{ref}$ (black dots) for a range of lines in measurement run (4) before correcting the FTS scale. A linear fit (red line) is used to determine $\lambda^{corr}_{ref}$. (b) The results of a $\lambda'_{ref}$ optimization scan performed after scale correction to confirm that the wavenumber dependence is reduced below the uncertainty. The solid and dashed red lines mark the final reference wavelength value (taken as the mean) and its estimated $1\sigma$ uncertainty range, respectively.